\documentclass[11pt,twoside]{article}
\usepackage{asp2010}

\resetcounters

\begin{document}

\title{NIRSPEC Radial Velocity Measurements of Late-M Dwarfs}
\author{A. Tanner$^1$, R. White$^1$, J. Bailey$^2$, T. Barman$^3$
\affil{$^1$Department of Physics and Astronomy, Georgia State University, P. O. Box 4106, Atlanta, GA 30316, USA}
\affil{$^2$Department of Astronomy, University of Michigan, 830 Dennison Bldg. 500 Church St. Ann Arbor, MI 48109-1042, USA}
\affil{$^3$Lowell Observatory, 1400 West Mars Hill Road, Flagstaff, AZ 86001, USA}}

\begin{abstract}

With an emphasis in detecting Earth-like planets set forth by the 2010 Decadal Survey and in searching for
planets around M dwarfs set forth by the 2008 Exoplanet Task Force, radial velocity surveys with infrared echelle spectrometers will have a significant impact on future exoplanet studies. Here, we present the results of an
infrared radial velocity survey of a sample of 14 late-M dwarfs with the NIRSPEC echelle spectrometer on the Keck II telescope. Using telluric lines for wavelength calibration, we are able to achieve measurement precisions
of 150-300 m/s over a year-long timeframe. While we require more RV epochs to determine whether most
of our stars have planetary-mass companions, we have placed upper limits of 5-10 M$_J$ on the masses of planets around a sub-set of our sample. We have also determined the rotational velocities for all the stars in the sample 
and offer our multi-order, high-resolution spectra over 2.0 to 2.4 $\micron$ to the modeling community to better understand
the atmospheres of late-M dwarfs. 
\end{abstract}

\section{Introduction}

Optical radial velocity surveys have resulted in the detection of hundreds of exoplanets with a
wide variety of characteristics including many surprises. Now, with measurement accuracies consistently at
the 1 m/s level with improvements in instrumentation expected to push that down to 0.1 m/s (Megevand et al. 2010), optical RV surveys
are capable of detecting super-Earth mass planets around some of our nearest M dwarf neighbors.  M dwarfs play a critical role in both of these scientific pursuits. The two competing planet formation paradigms, core accretion and gravitational instability, predict very different planet frequencies for the smallest stars; core accretion models predict relatively few planets (1\%; Kennedy \& Kenyon 2008) while gravitational instability models (e.g. Boss et al. 2006) predict frequencies similar to those of solar-type stars, perhaps 10\% or more. Despite M dwarfs contributing the majority of the planets with masses less than 10 M$_\oplus$sini,
they have been neglected in many of the original large scale RV programs since they are so optically faint (V $>$ 14) which
degrades the S/N of a given RV measurement and thus reduces the measurement accuracy. Finally, M dwarfs are, by far, the most abundant spectral type in the stellar neighborhood --- 73\% of all stars within 10 pc are M dwarfs (Henry et al. 2006). 

One solution to the faintness problem is to move the observations to the near-infrared part of the spectrum where M dwarfs
are considerably brighter than solar type stars. In addition, the contrast between the photosphere of the star and the starspots common among M dwarfs is smaller. These co-rotating features are a problematic source of noise when looking for the periodic RV signal produced by an orbiting planet. The periodicity of activity on the photosphere has been especially problematic when searching for planets around young stars (Setiawan et al. 2008; Huelamo et al. 2008; Prato et al. 2008).

\section{Observations and Analysis}

We have observed a small sample of 14 late M dwarfs (see Table~\ref{sample}) with the NIRSPEC spectrometer on the W. H. Keck II telescope (Mc Lean et al. 1998).  In total, the observation runs span over three years with some of the targets observed 2006/2007 and others observed in 2009. With this instrument we collected spectra in the NIRSPEC-7 (2.22 $\mu$m) passband in echelle mode with the 3 pixel slit (0.432$\arcsec$), at an echelle angle of 62.65 and grating angle of 35.50. This configuration results in a wavelength coverage of 1.99 to 2.39 $\micron$ and a  spectral resolution of approximately R=24000. While the echelle data has seven full spectral orders, for the radial velocity measurements, we focus on the 33rd order covering the spectral range of 2.28-2.32 $\micron$. This wavelength region is ideal for this purpose since it is permeated by telluric absorption features as well as the CO bandhead prominent in M dwarf spectra.

\begin{table}[!ht]
\caption{Sample of Late M dwarfs \label{sample}}
\smallskip
\begin{center}
{\small
\begin{tabular}{ccccc}
\tableline
\noalign{\smallskip}										
Target & RA [2000] & Dec [2000] & Spectral Type & K$_s$\\
\noalign{\smallskip}
\tableline
\noalign{\smallskip}	
2M0027+22	&	00 27 55.9	&	+22 19 32.8	&	M8.0	&	 9.57	         		\\
2M0140+27	&	01 40 02.6	&	+27 01 50.6	&	M8.5	&	11.43			\\
2M0253+16	&	02 53 00.9	&	+16 52 53.3	&	M6.5	&	7.59  			\\
2M0253+27	&	02 53 20.3	&	+27 13 33.2	&	M8.0	&	11.48			\\
2M0320+18	&	03 20 59.7	&	+18 54 23.3	&	M9.0	&	10.64			\\
2M1546+37	&	15 46 05.4	&	+37 49 45.8	&	M7.0	&	11.41			\\		
2M1835+32	&	18 35 37.9	&	+32 59 54.6	&	M8.5	&	9.17	        		\\
2M1707+64	&	17 07 18.3	&	+64 39 33.1	&	M9.0	&	11.38			\\
2M1757+70	&	17 57 15.4	&	+70 42 01.2	&	M7.5	&	10.40			\\
GJ 752 B       &      19 16 57.6       &      +05 09 01.6        &       M8.0 &       8.77          \\
2M2052-23	&	20 52 08.6	&	-23 18 09.6	&	M6.5	&	11.29			\\
2M2306-05	&	23 06 29.3	&	-05 02 28.6	&	M7.5	&	10.30			\\
2M2313+21	&	23 13 47.3	&	+21 17 29.4 	&	M6.0	&	10.44			\\
2M2235+18	&	22 35 49.1	&	+18 40 29.9	&	M7.0	&	11.37			\\
2M2349+12      &       23 49 49.0        &       +12 24 39.0          &      M8.0&        11.56                       \\
\noalign{\smallskip}
\tableline
\end{tabular}
}
\end{center}
\end{table}

The details of the data reduction and spectral extraction are given in Bailey et al. (2011 in preparation) and the full description of the results of this survey are given in Tanner et al. (2011 in preparation). Here, we summarize the data analysis and results of the survey. After flat fielding and dark correction, the 1-D spectra are extracted from the NIRSPEC images using an optimized method first proposed by Horne (1986). The wavelength
calibration  and initial values of the properties of the spectrograph are derived from a set of A star calibrators observed over the same night as the science targets. A stars are good calibrators in this instance since the spectrum of the star is featureless in the wavelength region used to estimate the radial velocities. For some of the spectra, we removed bad pixels by hand by replacing the significantly deviant pixels with the median value of the surrounding continuum. 

\begin{figure}[!ht]
\plotone{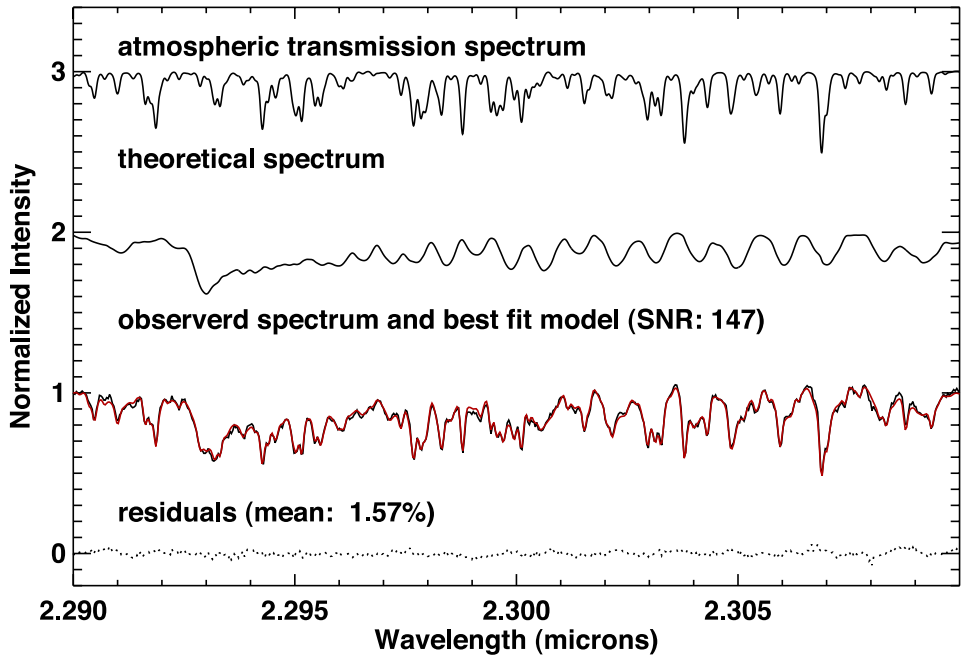}
\caption{Plot of one epoch of the telluric (top), synthetic model (middle) and combined, modeled (bottom-red) and observed (bottom-black) spectra used to determine  radial velocities for 2M1757+70. The bottom,
dotted line shows the residuals after the subtraction of the modeled and observed spectra which have an rms of $\sim$1.6\%. 
\label{fit}}
\end{figure}

The observed spectrum is the combination of the intrinsic spectrum of the star and the telluric spectrum convolved with the instrument 
point spread function (PSF). We utilize a telluric spectrum derived from a solar spectrum (Livingston \& Wallace 1991) and NextGen model stellar atmospheres of the late M stars derived as a function of effective temperature and surface gravity (Hauschildt et al. 1999).  For the A stars, we assume a flat, featureless intrinsic stellar spectrum. Figure~\ref{fit} shows the best fit to the data for the June 27, 2007 epoch.  Also, plotted are a model atmospheric transmission spectrum and the theoretical spectrum of an M8 dwarf. The bet fit to the data is determined through $\chi^2$ minimization. 

\section{Results}

\begin{figure}[!ht]
\plotone{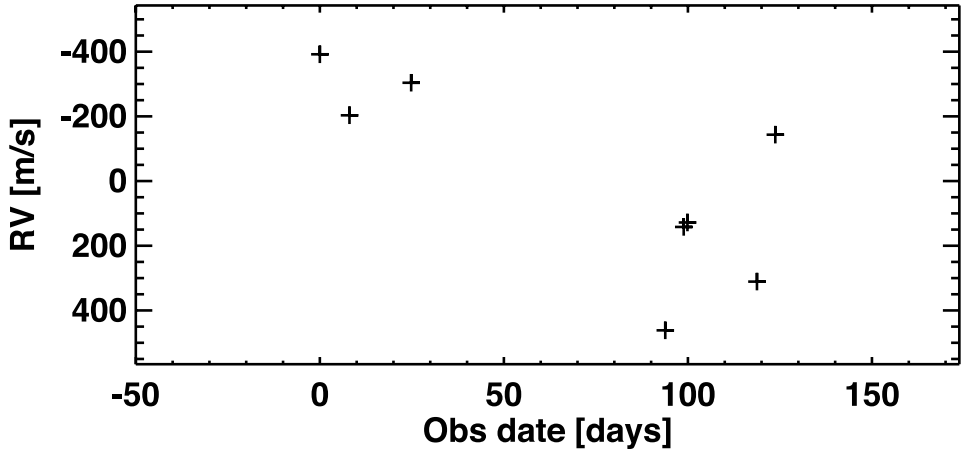}
\caption{Plot of the radial velocity measurements collected for GJ 752 B over a five months. These observations have an RMS of 250 m/s.\label{rvs}}
\end{figure}

Out of the 14 late-M dwarfs in our science sample, five of them  have multiple observations allowing us to
determine the rms of their nightly radial velocity measurements. For instance,  the late-type M dwarf, GJ 752 B has six epochs of RV 
measurements over the span of five months (see Figure~\ref{rvs}). These radial velocity measurements have an rms of 250 m/s. This is worse than the 50 m/s we have been getting for our early-M dwarf calibration stars (Bailey et al. 2010, in preparation) and the 100 m/s rms that we have estimated assuming a perfect instrument and no intrinsic stellar jitter (Butler et al. 1996). This could be due to the lower S/N of these observations or needed improvements in the synthetic model for the atmosphere of the M dwarf. 

 \subsection{Mass Limits Determined from RV Data}
 
Two of the stars, GJ 752B and 2M1757+70, have
six and seven epochs of RV data, respectively, over a time scale of more than a year. This allows us to constrain on the mass of any undetected planets orbiting these stars through Monte Carlo simulations.
Since this survey is most sensitive to short
period companions, we set limits only at representative orbital periods,
namely 10 and 100 days.  For each star, a set of 10,000 orbits are generated
with random inclinations, phases, and companion masses ranging from $\sim
1$ to $30$ Jupiter masses. Conservative detection limits are then set by determining the companion
mass that induces a RV dispersion that is 2$\sigma$ greater than the
expected dispersion for the star's $v$sin$i$ 99\% of the time.  With these assumptions, we can rule out planets around
these two stars with masses of more than $\sim$6 and 10 M$_J$ at periods of 10 and 100 days, respectively. 
 
\subsection{Rotational Velocities of Late-M dwarfs}

\begin{figure}[!ht]
\plotone{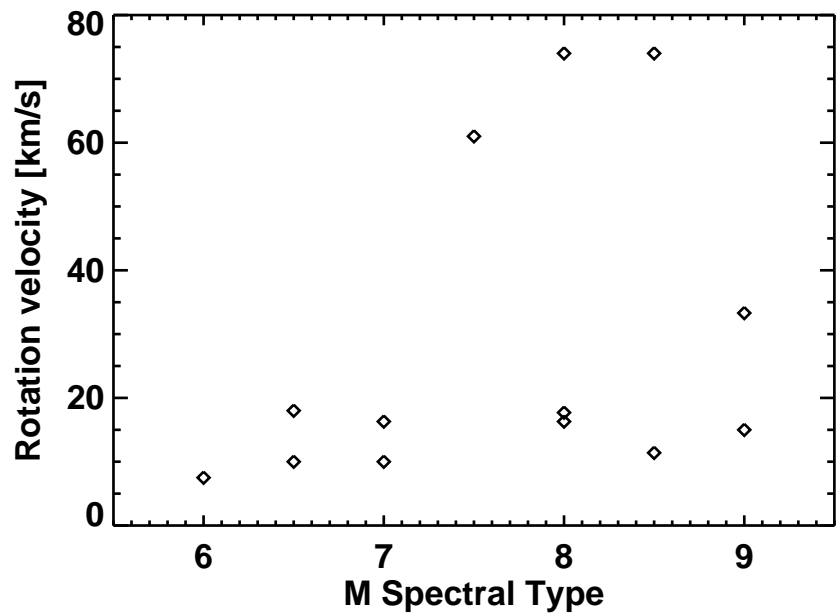}
\caption{Plot of our estimated rotational velocities as a function of stellar spectral type. These results resemble the larger sample of Reiners \& Basri (2009) which show no correlation. We do have evidence that
the precision of the RV measurements increases with rotational velocity.\label{vsini}}
\end{figure}

Because the rotational velocity is one of the free parameters used when modeling the observed spectrum, we
are able to determine this parameter for all of the stars in our sample. Figure~\ref{vsini} plots the rotational
velocities estimated from our infrared spectra as a function of spectral type. In some cases these values are similar to those
estimate from optical spectra (Reiners  \& Basri 2010) with good agreement between the infrared and optical
rotational velocities for the slow rotators ($<$20 m/s). However, the two fastest rotators in the sample have
disparaging infrared and optical rotational velocities. Reiners et al. 2009 and other have used starspot models
to show that the radial velocity jitter of a star is directly related its rotational velocity. Therefore, determining the rotational
velocities for a large portion of the thousands of M dwarfs within 25 pc and proving a rotational velocity versus radial 
velocity precision correlation observationally would greatly benefit future planet search efforts. 

\section{Summary and Recommendations}

With radial velocity precisions of 150-200 m/s for stars of masses of 0.08-0.15 M$_\odot$, in general, these observations are sensitive to planets with masses of 5-10 M$_J$ and 10-15 M$_J$ at periods of 10 and 100 days. Considering that there are a few thousand known M dwarfs within 25 parsec and these stars represent a population of stars largely unexplored with most planet detection techniques, these observations open the door to potentially fruitful new discoveries. 
While the percentage of early-M dwarfs with gas giant planets is currently estimated at 3\% (Johnson et al. 2010), we have no information on the same value for late-type
M dwarfs. With the masses of these stars approaching that of the most massive brown dwarfs and reaching a mass ratio of Y compared to a Jupiter mass planet, the planet population around these stars could have significant implications on the method of planet formation that is dominant in this stellar mass regime. 

\begin{figure}[!ht]
\plotone{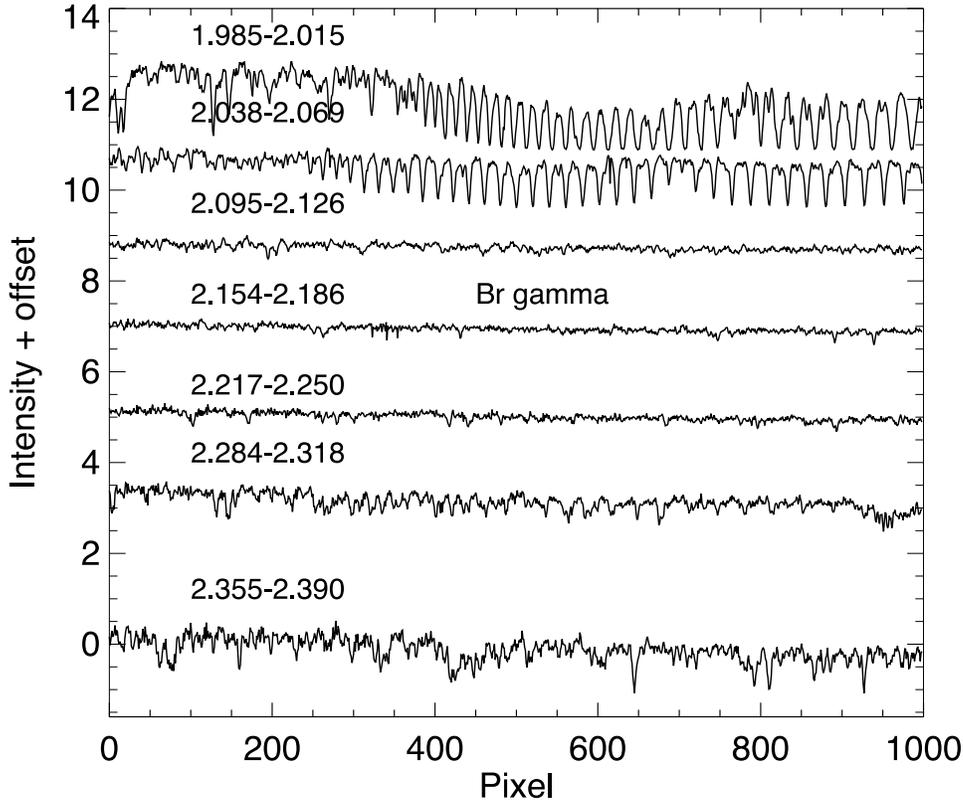}
\caption{Plot of all the observed spectral orders with our NIRSPEC setup of the M3.5 dwarf GJ 628. These spectra include both the CO bandhead and Br $\gamma$. \label{allorders}}
\end{figure}

These spectra can be used for more than just radial velocity measurements. Our data represent some of the only high-spectral resolution 
near-infrared spectra of late-M dwarfs published to date. While the radial velocity program uses just the 2.28-2.31 $\micron$ order, our NIRSPEC data cover 1.98-2.39 $\micron$ (see Figure~\ref{allorders}). With the discovery of brown dwarfs a few decades ago, modeling the spectral features and, therefore, the complex atmospheres of late-M dwarfs became overlooked compared to what's be accomplished for solar-type stars. Our M dwarfs spectra are available to any who wish
to utilize them for spectral synthesis. None of our spectral show signs of Br $\gamma$ emission (2.166 $\micron$) which could be used to aid in age determinations.  

With so few targets in this pilot study, its difficult to determine whether there is any observational correlation between radial velocity  precision and the rotational velocity or spectral type of the star. However, we have proven that we are capable of detecting Jupiter-mass planets around these stars which are too faint to be
observed efficiently with optical echelle instruments. Planned upgrades to the NIRSPEC instrument which include adding a methane gas absorption
cell to improve wavelength calibration could lead to radial velocity precisions of $\sim$20-30 m/s (Bean et al. 2009). Such observations would be sensitive to Jupiter mass planets in orbits of less than a year around late-M dwarfs.


\begin{references}
\reference{} Bean, J., et al., 2007 AJ, 134,  749
\reference{}  Bean, J. et al. 2009 ApJL, 711, L19
\reference{} Greene, T. P., Tokunaga, A. T., Toomey, D. W., \& Carr, J. B. 1993, in SPIE Conference Series, ed. A. M. Fowler, Vol. 1946, 313Ð324
\reference{} Johnson, J.~A., Aller, K.~M., Howard, A.~W., \& Crepp, J.~R.\ 2010, PASP, 122, 905 
\reference{} Hauschildt, P.~H., Allard, F., Ferguson, J., Baron, E., \& Alexander, D.~R.\ 1999, ApJ, 525, 871 
\reference{} Henry, T.J., Jao, W.-C.,  Subasavage, J.P., Beaulieu, T.D., Ianna, P.A., Costa, E.,  \& Mendez, R.A., 2006, AJ, 132, 2360  
\reference{} Hinkle, K. H., et al. 2003, in SPIE Conference Series, ed. P. Guhathakurta, Vol. 4834, 353Ð363
\reference{} Horne, K.\ 1986, PASP, 98, 609 
\reference{} Hu{\'e}lamo, N., et al.\ 2008, A\&A, 489, L9
\reference{} Kaufl, H.-U., et al. 2004, in SPIE Conference Series, ed. A. F. M. Moorwood \& M. Iye, Vol. 5492, 1218Ð1227
\reference{} Livingston, W., \& Wallace, L.\ 1991, NSO Technical Report, Tucson: National Solar Observatory, National Optical Astronomy Observatory, 1991,  
\reference{} Kennedy, G.~M., \& Kenyon, S.~J.\ 2008, \apj, 673, 502
\reference{} McLean, I.~S., et al.\  1998, SPIE, 3354, 566 
\reference{} Megevand, D., et al.\ 2010, SPIE, 7735,165
\reference{} Prato, L., Huerta, M., Johns-Krull, C.~M., Mahmud, N., Jaffe, D.~T., \& Hartigan, P.\ 2008, \apjl, 687, L103 
\reference{} Pravdo, S.~H., \& Shaklan, S.~B.\ 2009, ApJ, 700, 623 
\reference{} Reiners, A., \& Basri, G.\ 2010, ApJ 710, 924 
\reference{} Setiawan, J., Henning,  T., Launhardt, R., M{\"u}ller, A., Weise, P., Kurster, M.\ 2008, Nature, 451, 38 
\end{references}
\end{document}